\documentclass[12pt,twoside]{article}
\usepackage{xrb2007}
\usepackage{natbib, graphicx}
\begin{document}

% select your session by uncommenting the appropriate line
\session{Jets}
%\session{Jet and Black Hole Binaries}
%\session{Faint Galactic XRB Populations}
%\session{Faint XRBs and Galactic LMXBs}
\session{Obscured XRBs and INTEGRAL Sources}
%\session{ULXs}
%\session{Extragalactic Populations}
%\session{Future Missions and Surveys}
%\session{Population Synthesis}

\shortauthor{Winkler}
\shorttitle{INTEGRAL and New HMXB Classes}

\title{INTEGRAL and New Classes of High-Mass X-Ray Binaries}
\author{Christoph Winkler}
\affil{Astrophysics Division, Research and Scientific Support Department, ESA-ESTEC, Keplerlaan 1,
2201 AZ Noordwijk, The Netherlands}

\begin{abstract}
The gamma-ray observatory INTEGRAL, launched in October 2002, produces
a wealth of discoveries and new results on compact high energy 
Galactic objects, nuclear gamma-ray line emission, diffuse line and continuum emission,
cosmic background radiation, AGN and high energy transients.
Two important serendipitous discoveries made by the INTEGRAL mission are 
new classes of X-ray binaries, namely the highly-obscured high-mass X-ray 
binaries, and the super-giant fast transients.
In this paper I will review the current status of these discoveries. 
\end{abstract}

\section{Introduction}
The ESA gamma-ray observatory INTEGRAL \citep{winkler03} is dedicated to the 
fine  spectroscopy (2.5 keV FWHM @ 1 MeV) and fine
imaging (angular resolution: 12 arcmin FWHM) of celestial gamma-ray 
sources in the energy range 15 keV to 10 MeV with concurrent source 
monitoring in the X-ray (3-35 keV) and optical (V-band, 550 nm) bands.
While the pre-planned scientific observing programme of the mission is driven by the usual 
Announcement of Opportunities with peer reviewed observing proposals, important
\underline{\em{serendipitous discoveries}} have also been made by INTEGRAL. Two of them are 
briefly described in this paper. Ingredients for new discoveries of point sources
above 15 keV are: (i) a very large field of view of almost 900 square degrees, 
(ii) arc-minute source location capability, and
(iii) a broad energy band with good sensitivity. In addition, long exposures
and/or frequent monitoring of the inner Galaxy are provided through the general
open time observing programme, the core programme
(\cite{gehrels}, \cite{winkler01}) and via the key programmes
introduced recently. Nature, finally, contributes the highly variable
high-energy sky. Monitoring the sky in the INTEGRAL energy range is of
fundamental importance for Targets of Opportunities, multi-wavelengths follow-up
observations and serendipitous discoveries.
\section{HMXB and New IGR Sources}
Recent hard X-ray sky surveys with INTEGRAL (\cite{bird}, \cite{bodaghee}, \cite{krivonos}) 
have produced source catalogues containing between 400 and
500 detected point sources, covering energy intervals between 15 keV to 100 keV.
These surveys differ
slightly in sky  coverage, energy range, time interval, detection criteria and
detection software, but we can conclude that, on average, about 17\% of all 
detected sources are HMXB (73 sources, on average) out of which one third are 
new INTEGRAL gamma-ray (IGR) sources identified
as HMXB. About 23\% of all
detected sources remain yet to be identified. 
In the following sections we will
briefly describe the new HMXB classes as detected by INTEGRAL: highly obscured
HMXB and super-giant fast X-ray transients. Both classes are populated by binary
systems, where the compact object is orbiting a massive early-type OB
super-giant star. These systems contribute to 
about 20\% of the Galactic HMXB population, while the remaining 
80\% of HMXB are binaries involving a compact object orbiting a Be star.

\subsection{Highly Obscured HMXB}
\begin{figure*}[htp]
   \centering
   \includegraphics[width=9 cm]{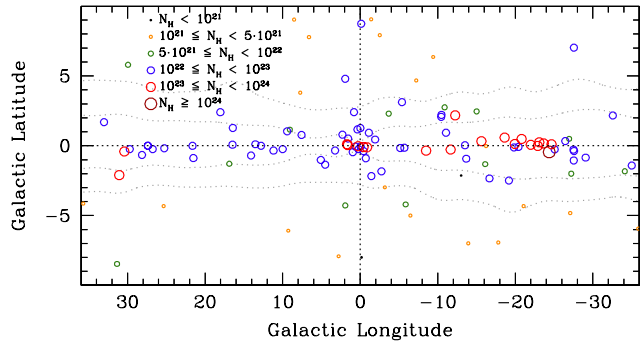}
      \caption{Spatial distribution (inner Galaxy) of all galactic sources
      (mostly HMXB) detected by INTEGRAL, for which N$_H$ has been reported.
      From \citet{bodaghee}. 
}
         \label{hmxb}
   \end{figure*}
 \vfil
IGR J16318$-$4848 was discovered by INTEGRAL on 29 
January 2003 \citep{courvoisier}, shortly after the start of the nominal
observing programme. The source is located very close to the Galactic 
plane (30$^\prime$
off), and it was found that strong absorption below 5 keV (N$_H$
$\sim$ 2$\times$10$^{24}$ cm$^{-2}$) is dominating  the
spectral distribution of the HMXB with a compact object enshrouded in a
Compton
thick environment \citep{walter}. Subsequent long term monitoring 
confirmed that
the source remained bright and Compton thick \citep{ibarra}. Using data from
XMM-Newton, ASCA and RXTE, more INTEGRAL detected 
IGR sources with similar broad band ''highly
absorbed'' spectra could be identified \citep{lutovinov}. How 
many of these
sources have been detected (until end 2006) by INTEGRAL? 
The recent survey by
\citet{bodaghee} displays, as shown in Fig.~\ref{hmxb}, the spatial distribution of all 
Galactic
sources (mostly HMXB) detected by the imager IBIS on-board 
INTEGRAL for which N$_H$ has been reported. If we define ''highly
absorbed'' as local absorption in excess of ISM absorption in the line-of-sight, that is
N$_H$ $>$ 10$^{23}$ cm$^{-2}$, then we conclude from Table 1 in \citet{bodaghee}
that out of 25 HMXB meeting this criterion, 16 sources (64\%) are new IGR/HMXB
sources, and 9 sources (36\%) have been known previously. 
The ''typical'' source geometry and population characteristics 
are as follows: (i) a compact source embedded in dense material; (ii) 
the fluorescence region is larger than the orbital radius; (iii) spherical
geometry; (iv) unknown or weakly detected in X-ray surveys prior to INTEGRAL;
(v) strong low energy absorption N$_H$ $>$ 10$^{23}$ cm$^{-2}$; (vi)
predominantly located in the Galactic bulge and along the Norma/Scutum spiral arms; (vii)
long spin periods (typically 100 s to 1300 s) characteristic of wind accretion; (viii)
short orbital periods $<$ 10 days; (ix) early type stellar super-giant
companion (\cite{walter05}, \cite{bodaghee}). In summary, the absorbed
systems $-$ a new class of HMXB $-$ form the majority of active super-giant
HMXB in our Galaxy. Thanks to the large increase of known HMXB in the inner
Galaxy, it is now possible to study the HMXB spatial distribution in the
inner Galaxy and to compare it with the location of star forming regions and spiral arm
patterns (e.g. \cite{lutovinov}, \cite{bodaghee}).

\subsection{Super-Giant Fast X-Ray Transients}
Thanks to the ''ingredients'' mentioned in the introduction, 
another (sub)class of HMXB previously hidden
throughout the Galaxy could be identified with INTEGRAL: the super-giant fast X-ray transients (SFXT).
While X-ray transients in general exhibit variations on timescales from few days to weeks or months, the
SFXT are characterized by short outbursts lasting typically up to 
a few hours. Before INTEGRAL, the nature of
these transients was largely unknown (see \cite{sguera} for a summary). With INTEGRAL  
it was possible to
detect new SFXT; detect for the first time recurrent outbursts, and associate SFXT 
with super-giant HMXB known to be persistent sources \citep{sguera06}. 
SFXT form a new sub-class of super-giant HMXB and
we can assume that there are many more SFXT in our Galaxy as previously thought.
The origin of the fast outburst, during which the persistent luminosity 
of $\sim$10$^{32}$ erg/s increases by
about 3 orders of magnitude is not firmly established: sudden accretion of small ejections originating
in a clumpy wind, outbursts near or at periastron, or due to a second wind component (equatorial disk)
of the super-giant donor (see \cite{sidoli} for a discussion). Up to now (October 2007), eight firm
detections of 
SFXT\footnote{IGR J08408$-$4503, IGR J11215$-$5952, IGR J16465$-$4507, XTE J17391$-$3021,
IGR J17544$-$2619, SAX J1818.6$-$1703, AX J1841.0$-$0535, and IGR J18450$-$0435}
exist with  20 more candidates under current investigation (\cite{sguera06}, \cite{bird},
\cite{bodaghee}, \cite{bazzano}). 
\section{Conclusions}
INTEGRAL discovered new classes of highly absorbed HMXB and SFXT, thanks to its large FOV combined with
a broad energy range, fine imaging, good sensitivity and observing strategy. Long term monitoring with a
large FOV is crucial to further enlarge the database. The new discoveries also
pose new problems to be addressed by future observations: Why do we  find highly absorbed HMXB as 
slowly rotating neutron stars with modest magnetic fields only, but no cyclotron lines and no 
black hole candidates? How can we explain the fast luminosity increase of SFXT 
by about 3 orders of magnitude?

\end{document}